\documentclass[sigplan, screen]{acmart}

\setcopyright{none}
\acmConference[SOAP '26]{15th ACM SIGPLAN International Workshop on the State Of the Art in Program Analysis}{June 16, 2026}{Boulder, CO, USA}
\acmYear{2026}
\acmDOI{}
\acmISBN{}

\usepackage{tabularx}
\usepackage{tikz} 
\usetikzlibrary{shapes.geometric, arrows}
\usepackage{balance}

\setcitestyle{numbers, square}


\renewcommand\footnotetextcopyrightpermission[1]{}

\begin{document}

\title{Conditional Execution of Transpiler Passes Based on Per-Script Feature Detection}

\author{Rishipal Singh Bhatia}
\affiliation{%
  \institution{Google}
  \city{Sunnyvale}
  \state{California}
  \country{USA}
}
\thanks{Preprint. Under review at the 15th ACM SIGPLAN International Workshop on the State Of the Art in Program Analysis (SOAP '26).}
\email{rishipal@google.com}

\begin{abstract}
As the ECMAScript specification evolves, industrial-scale JavaScript compilers face the challenge of supporting modern language syntax while maintaining compatibility for diverse execution environments. Traditionally, compilers solve this by running transpilation passes in a monolithic pipeline, where the transpilation passes are chosen to execute strictly based on a target language level. This results in significant computational waste, as compilers perform expensive Abstract Syntax Tree (AST) traversals to lower features that may not exist in the actual input source code. We present a compiler improvement that conditionally executes transpiler passes based on accurately tracking and dynamically maintaining the exact set of language features present in the compilation unit throughout the transpilation process. It is implemented in the production Google Closure Compiler. By populating and maintaining a FeatureSet at every JavaScript script-level, it dynamically skips running unnecessary lowering passes. We detail the architectural safeguards---including strategic pass ordering and dynamic validation of the transpiled code for feature-correctness. Evaluation of this improvement on large-scale production monorepos produced a considerable reduction in compilation time and saved compute and memory usage.
\end{abstract}


\maketitle

\section{Introduction \& Problem Statement}
JavaScript development has shifted toward modern language syntax (ECMAScript 2022+), driven by the rapid evolution of the ECMAScript specification \cite{1}. However, production deployment remains constrained by a long tail of legacy environments that require ECMAScript 5 or ECMAScript 6 output levels. This includes older browser engines, specialized embedded runtimes, and non-browser environments. This gap is bridged by transpilation: the source-to-source lowering of modern language constructs into functionally equivalent but more compatible patterns.

In large-scale industrial web applications, such a transpilation step is a high-frequency, resource-intensive operation within the build environment. Google Closure Compiler \cite{2}, which powers some of the world's largest JavaScript applications, includes transpilation as a core phase of the compilation process. Historically, the compiler utilized a Range-Based Execution Model. In this model, the selection of transpilation passes was determined solely by the distance between the input language version and the requested output language version. For example, if a compiler was tasked with producing ES5 output, it would automatically queue the entire sequence of lowering passes spanning from the current ECMAScript version (ESNext) down to ES5.

The fundamental flaw in this approach is its lack of granularity regarding the actual content of the source code. Most compilation units do not utilize the full breadth of modern language features. For instance, a script might use the ES2020 optional chaining feature, but contain no ES2015 classes. Despite this, the range-based model would force the compiler to execute the set of class-lowering passes, incurring the full overhead of pass setup and AST traversal. In a massive monorepo comprising millions of files, these accumulated redundant traversals create a significant bottleneck, inflating build latency and consuming vast amounts of unnecessary CPU cycles.

\section{Background: The Transpiler Pipeline}
To understand why the redundant traversals are so costly, it is necessary to examine the architecture of the Google Closure Compiler’s transpiler.

\subsection{Incremental Lowering \& Pass Ordering}
The Google Closure Compiler decomposes transpilation into a series of tiered, atomic transformation passes that run serially on scripts; one script at a time. Each pass is responsible for lowering specific features from a newer language level to an older language level on the script it is lowering. For example:
\begin{itemize}
    \item \textbf{RewriteOptionalChainingOperator pass:} rewrites the ECMAScript 2020 optional chaining syntax (\texttt{?.}) operator with conditional syntax (\texttt{? :}).
    \item \textbf{Es7RewriteExponentialOperator pass:} rewrites the exponential operator syntax of ECMAScript 7 into ECMAScript 5 \texttt{Math.pow} operations.
    \item \textbf{Es6RewriteRestAndSpread pass:} rewrites the ECMAScript 6 rest and spread syntax into polyfills containing ECMAScript 5 syntax.
\end{itemize}

These passes are executed in a strict reverse-chronological order of features based on the language specification year. For example:
\begin{itemize}
    \item \textbf{Cascading Transformation:} High-level features like “async/await” (an ECMAScript 2017 feature) are first lowered into ECMAScript 2015 (ES6) generators (\texttt{funct\allowbreak ion*}, \texttt{yield}). 
    \item \textbf{Sequential Reduction:} Only after the async transformation is complete can a subsequent pass lower the newly created Generators into ES5-compatible state machines.
\end{itemize}

\subsection{The Cost of a Transpiler Pass}
In the legacy workflow, once the compiler identified the output language level, it would instantiate and run every pass required to reach that target level. Each active pass involves:
\begin{enumerate}
    \item \textbf{Context Initialization:} Setting up internal state required for the pass.
    \item \textbf{Full AST Traversal:} A comprehensive "walk" of the entire tree to look for relevant nodes.
    \item \textbf{Node Rewriting:} Performing the actual transformation if a feature is found.
    \item \textbf{Cleanup \& Updates:} Updating the compiler state to record that the rewriting is complete, and ensuring AST integrity after the pass.
\end{enumerate}
The primary inefficiency is that steps 1, 2, and 4 are performed for every pass in the range, regardless of whether step 3 (the actual work) ever occurs.

\subsection{Downward Monotonicity of Language Levels}
The architectural integrity of the transpiler relies on the assumption of downward monotonicity. While the ECMAScript specification organizes features by year (language level $L_N$), the transpiler implementation utilizes a set of fine-grained, atomic passes $P_{f, N}$ for lowering the individual language features $f \in L_N$.

Formally, each pass $P_{f, N}$ functions as a transformation $T: f \rightarrow \{f' \in L_k \mid k < N\}$. The subsequent chain of passes continues this reduction iteratively toward the minimum target level. This structure relies on the downward monotonicity assumption: for any pass $P_f$ targeting feature $f$, no subsequent pass $P_{f'}$ in the pipeline can reintroduce $f$ into the AST. 

The solution described in the next section enhances the feature tracking during transpilation, which allows the compiler to convert that monotonicity assumption into an enforced invariant during compilation. It leverages this granularity of transformations and enhanced feature tracking to unlock the ability to skip a specific subset of passes within a single language level, most notably within the expansive ES6 $\rightarrow$ ES5 lowering chain. By accurately detecting feature presence at the script level, we can prune these atomic traversals even when other features from the same specification year are present.

\section{Improvement: Selective Transpilation}
The improvement presented in this paper is implemented in the Google Closure Compiler as a “Selective Transpilation” model \cite{4, 9}. This approach introduces an analytical "gate" that protects the compiler from executing redundant work by ensuring that lowering passes are only invoked when their respective language features are present within a specific script's Abstract Syntax Tree. 

\subsection{Recording Input Features}
As a first step, the compiler recognizes and stores the set of features encountered by the parser in the input scripts. The parser in Google Closure Compiler identifies usages of specific language constructs in a given input script, and records them as features belonging to that script. This recording is done as an attribute on every \texttt{SCRIPT} node, which is the top level AST node in the compiler’s internal AST representation for the file. To avoid increasing any memory pressure, the parser operates purely on the existing AST structure without creating temporary objects or secondary data structures beyond the node attribute. 

We formalize the set of all language features recognized by the compiler as $F$. For a script $S$, we define the script feature set $\Phi(S) \subseteq F$ such that:
\[ \Phi(S) = \{ f \in F \mid \exists n \in S : \text{isFeatureNode}(n, f) \} \]

\subsection{Mapping transpiler passes to their features}
Each transpiler pass cleanly defines the set of features that it is expected to lower $\psi(P) \rightarrow f$, and exposes that API to the compiler. This allows the compiler to retrieve the mapping of the transpiler pass to the relevant features that must trigger it. 

\subsection{Accurately Tracking the FeatureSet}
A critical technical challenge in selective omission is ensuring the FeatureSet remains accurate across the entire transpilation process. If a transpiler pass is skipped from running on a script based on an outdated FeatureSet, or if a pass introduces a feature that goes untracked (i.e. fails to register the new feature in the script’s FeatureSet), the compiler may fail to lower required syntax.

\subsubsection{Synthetic Feature Propagation}
Features are conceptually of two types: Source Features (detected in the input source code during parsing) and Synthetic Features (introduced by transpiler passes during lowering).

The FeatureSet is a dynamic property. So, when a pass $P_i$ lowers a high-level feature $f_{source}$ (e.g., converting an \texttt{AsyncFunction} to a \texttt{Generator}), it must programmatically notify the compiler of the new features it has introduced. The compiler then performs an in-memory update: $\Phi(S) = (\Phi(S) \setminus \{ f_{source} \}) \cup \{ f_{synthetic} \}$.

This update happens immediately upon the completion of the pass, ensuring the scheduler has an up-to-date FeatureSet on every script before evaluating the skip-predicate for the next pass, $P_{i+1}$. Without this propagation, a pass targeting the synthetic feature (e.g. Generators) might be erroneously skipped if the original source code did not contain them, causing a compile-time error at the end of transpilation.

\subsubsection{The Transpiled-Away Set}
To prevent silent failures, the compiler maintains a global Transpiled-Away Set ($T$) throughout transpilation. This set records every feature $f$ that has been transpiled-away by its corresponding transpilation pass. Once a pass $P_i$ completes executing over all scripts in the compilation unit, feature $f$ is added to $T$.

The system enforces a strict invariant: Any feature in $T$ must not exist in $\Phi(S)$ for any script S in the compilation unit. If a pass $P_j$ (where $j > i$) attempts to introduce a feature $f$ that is already in $T$, the compiler flags a compile-time violation. This ensures that the cascading nature of transpilation is strictly one-way and that features are never re-introduced after their specific lowering pass has retired.

\subsubsection{Transpiler Pass Controlled FeatureSet}
The FeatureSet is stored as a persistent attribute of the \texttt{Script} node in the AST. Transpiler passes that manipulate the tree are required to maintain this set. To minimize overhead, the compiler does not separately re-scan the AST after every rewriting pass in production mode to update the \texttt{FeatureSet}. Instead, it relies on the transpilation passes to "own" the FeatureSet property related to their specific features, with the compiler only validating the accuracy of the feature set at the end of transpilation. This "ownership" model ensures high performance while maintaining the accuracy required for selective omission.

\subsection{Selective Execution Logic}
During the execution of the compiler's main loop, the execution condition for a transpiler pass $P$ on a script $S$ is satisfied if there exists at least one feature handled by $P$ that is both unsupported by the target level and present in the script's feature set.

This is formalized as: 
\[ \exists f \in \psi(P) : (f \notin \text{Features}(\text{TargetLevel})) \land (f \in \Phi(S)) \]

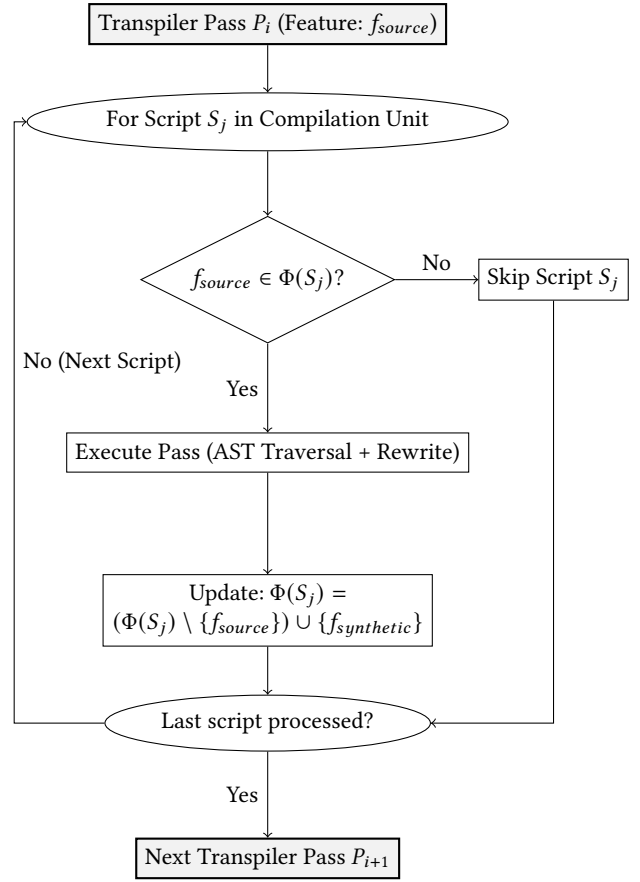
\begin{figure}[ht] 
\centering
\begin{tikzpicture}[node distance=1.8cm, font=\small]
\node (pass) [rectangle, draw, fill=gray!10, thick] {Transpiler Pass $P_{i}$ (Feature: $f_{source}$)};
\node (startloop) [ellipse, draw, below of=pass, yshift=0.5cm] {For Script $S_{j}$ in Compilation Unit};
\node (decide) [diamond, draw, aspect=2, below of=startloop, yshift=-0.3cm] {$f_{source} \in \Phi(S_{j})$?};
\node (run) [rectangle, draw, below of=decide, yshift=-0.5cm] {Execute Pass (AST Traversal + Rewrite)};
\node (update) [rectangle, draw, below of=run, yshift=-0.3cm, align=center] {Update: $\Phi(S_{j}) =$\\ $(\Phi(S_{j}) \setminus \{ f_{source} \}) \cup \{ f_{synthetic} \}$};
\node (skip) [rectangle, draw, right of=decide, xshift=2cm] {Skip Script $S_{j}$};
\node (endloop) [ellipse, draw, below of=update, yshift=0.3cm] {Last script processed?};
\node (nextpass) [rectangle, draw, fill=gray!10, thick, below of=endloop] {Next Transpiler Pass $P_{i+1}$};

\draw [->] (pass) -- (startloop);
\draw [->] (startloop) -- (decide);
\draw [->] (decide) -- node[anchor=east] {Yes} (run);
\draw [->] (decide) -- node[anchor=south] {No} (skip);
\draw [->] (run) -- (update);
\draw [->] (update) -- (endloop);
\draw [->] (skip) |- (endloop);
\draw [->] (endloop) -- node[anchor=east] {Yes} (nextpass);

\draw [->] (endloop.west) -- ++(-1.2,0) |- node[anchor=west, pos=0.30] {No (Next Script)} (startloop.west);

\end{tikzpicture}
\Description{A flowchart showing the selective transpilation logic. It starts with a transpiler pass targeting a specific feature. It loops through each script, checking if the feature is in the script's feature set. If yes, it executes the pass and updates the feature set. If no, it skips the script. After all scripts are processed, it moves to the next pass.}
\caption{Selective Transpilation Logic: The compiler iterates through scripts for each pass, skipping scripts that lack the targeted feature $f_{source}$ according to their specific FeatureSet $\Phi(S)$.}
\label{fig:gate}
\end{figure}

We visualize this gating mechanism and its integration into the compiler's execution loop in Figure \ref{fig:gate}. The flowchart details how the compiler evaluates the presence of the targeted feature $f_{source}$ for each script individually, enabling the dynamic omission of the entire transformation pass for scripts that do not require it, while ensuring synthetic features are propagated correctly for subsequent passes.

\section{Architectural Safeguards}
Pruning transpiler passes dynamically in the compiler creates the risk of "feature leakage." We implement four tiers of safeguards to ensure soundness:
\begin{itemize}
    \item \textbf{Strategic Pass Ordering:} As described in Section 3, passes run in reverse chronological order. This ensures that lowered code does not re-introduce features from a higher tier.
    \item \textbf{Dynamic Feature Tracking:} The script nodes’ \texttt{Node.\allowbreak FEATURE\_\allowbreak SET} property is updated in real-time. If a transformation introduces a new feature that requires a later lowering pass, that feature is added to the set.
    \item \textbf{AST Validation:} An \texttt{ASTValidator} is injected during development time after every pass. It cross-references the current AST against a "Transpiled-Away" set. If a pass attempts to introduce a feature that has already been marked as removed, the compiler reports an error. Furthermore, if a pass attempts to introduce a permitted feature without explicitly reporting the change to the compiler, that gets caught as a compiler error too.
    \item \textbf{Post-Transpile Verification:} A production pass \texttt{post\-Transpile\-Check\-Un\-supported\-Features\-Re\-moved} runs at the end of the transpilation stage to guarantee that no unsupported features leaked through.
\end{itemize}

\section{Evaluation \& Results}
\subsection{Evaluation Methodology}
To provide a credible and reproducible assessment of the solution, we conducted our evaluation using a representative sample of large-scale production applications from the internal monorepo. These applications span several domains, including high-traffic search services, geospatial mapping tools, and complex web-based productivity suites. This diversity ensures the results account for various coding patterns, from legacy ES5 to modern ES2022 JavaScript.

The evaluation was performed in a standardized distributed build environment using cloud-based workers. Each worker was a Linux VM configured with consistent specs (e.g. a 32-core high-performance CPU and 64GB of RAM). To mitigate external noise, each benchmark was executed 3 times, and the mean values were recorded.

\subsection{Evaluation Results} 
\begin{itemize}
\item \textbf{Compile Time Improvement:} For select large-scale production web applications compiled via the Google Closure Compiler, the total transpilation time is reduced by 50\%. We observed a mean reduction in compilation time of 12\% in the specific stage of compilation (that contains transpilation) of all production web applications via the Google Closure Compiler \cite{3}. 

\item \textbf{Compute Savings:} The stage of compilation (containing the transpilation step) that was improved by this solution executes approximately 900,000 times every day for compiling Google’s production web applications. 

\item \textbf{Redundant Pass Analysis:} For a typical production web application targeting the ES5 output level, 30\%--40\% of transpilation passes were found to be performing redundant traversals in the legacy model. 

\item \textbf{Scalability:} As modern features continue to get introduced in the input JavaScript language (e.g. consider a hypothetical ES2027 feature set), the legacy model would continue to execute the entire range of transpilation passes (ES2027 $\rightarrow$ ES5) even if modern language features are sparsely used in the input source code. With the new model, the compilation cost is decoupled from the new, unused features.
\end{itemize}

\begin{table}[ht]
\caption{Impact of Selective Transpilation on Production Workloads}
\label{tab:results}
\small 
\setlength{\tabcolsep}{3pt} 
\begin{tabularx}{\columnwidth}{@{} X r @{}}
\toprule
\textbf{Metric} & \textbf{Result} \\ 
\midrule
Transpilation Time (Large Apps) & 50\% Reduction \\
Mean Compilation Stage Latency & 12\% Reduction \\
Redundant Transpiler Passes Skipped & 30\% -- 40\% \\
Daily Production Executions & $\approx$900,000 \\
\bottomrule
\end{tabularx}
\end{table}

\subsection{Implementation in Open Source} 
The selective transpilation model is available and enabled by default in the open-source distribution of the Google Closure Compiler. The central gating logic is encapsulated within the \texttt{doesScriptHaveAnyOfTheseFeatures} method in the \texttt{TranspilationPasses.java} utility class. 

For benchmarking or debugging purposes, this optimization can be toggled by modifying the aforementioned method to return a constant \texttt{true}. This effectively reverts the compiler to the legacy range-based execution model, forcing the execution of all available transpiler passes within the range on every script regardless of content. In practice, disabling this feature when building complex web projects results in a measurable increase in build latency due to the reintroduced redundant traversals.

\subsection{Verification of Correctness}
A critical requirement for this optimization was that it remained semantically transparent; skipping redundant passes must not result in different output code produced by the compiler. We employed several testing methodologies to verify that the Selective Transpilation model produced identical results to the legacy Range-Based model:
\begin{itemize}
    \item \textbf{Differential Testing:} For select internal applications, we compiled the source code using both the legacy and selective transpilation models. We then performed a bitwise comparison of the final minified production bundles to ensure they match exactly.
    \item \textbf{Code Size Testing:} We ran the code size regressions test suite to validate that skipping redundant passes (selective transpilation) makes no difference to the final output code size produced by the Google Closure Compiler. 
    \item \textbf{Compiler Test Suite:} We ran the existing Google Closure Compiler test suite (comprising thousands of unit tests) with the improvement enabled.
\end{itemize}

\section{Comparison with Related Work}
\textbf{Babel \& SWC:} Most popular transpilers \cite{5, 6} utilize a plugin-based architecture where users explicitly enable transformations. While SWC \cite{6} offers extreme speed through its Rust-based implementation, it still generally follows a range-based execution for configured presets. Selective Transpilation differs by automating this selection based on deep AST inspection rather than static configuration.

\textbf{LLVM Pass Manager:} LLVM \cite{7} employs a PassManager that can skip certain optimization passes based on analysis results. We extend this concept to high-level language features in a source-to-source context.

\textbf{Just-In-Time (JIT) Compilers:} Engines like V8 \cite{8} use dynamic profiling to optimize "hot" code paths. In contrast, Selective Transpilation is a static optimization that benefits the developer's build-time experience across the entire codebase.

\section{Future Work}
The success of Selective Transpilation suggests a broader pattern for compiler design. Possible extensions include:
\begin{itemize}
    \item \textbf{Selective Optimization:} Applying similar feature-based gating to expensive optimization passes, we can skip high-complexity optimizations if the language constructs they’re optimizing do not exist in the AST. \textbf{The Caveat:} Unlike transpilation, in which each transpiler pass has clear chronological spec boundaries, updating a "FeatureSet" to accurately reflect the AST state at the end of an optimization pass is significantly more complex.
    \item \textbf{Fine-Grained Scoped Analysis:} Performing the feature scan at the function level rather than the script level to allow passes to skip specific sub-trees of the AST. \textbf{The Trade-off:} While this could prune more work in massive "monolith" files, the overhead of storing and querying per-scope features would significantly increase memory consumption and complexity. For many average-sized files, the cost of the more granular analysis might exceed the time saved in the lowering passes.
\end{itemize}

\balance

\section{Conclusion}
The selective transpilation model described in this paper represents a significant architectural shift in how industrial-scale compilers manage the ever-expanding ECMAScript specification. By moving from a rigid, range-based execution model to a dynamic, feature-gated approach, we have successfully decoupled the computational cost of transpilation from the presence of unused modern language features. Our implementation in the Google Closure Compiler demonstrates that a lightweight static analysis, when combined with robust feature tracking and safeguards, can reclaim substantial build-time resources without compromising the semantic integrity of the output.

The observed reductions in compilation latency and CPU cycles confirm that the overhead of maintaining a dynamic FeatureSet at the script level is negligible compared to the time savings realized by skipping redundant AST traversals. Furthermore, the new model provides a sustainable path for compiler development; as the TC39 committee \cite{10} continues to introduce new syntax, the compiler's performance will remain proportional to the features actually utilized by the developers, rather than the total number of features defined in the standard.

Ultimately, the methodology described in this paper provides a blueprint for modern compilers to achieve high-performance source-to-source transformations in an era of rapid language evolution. By strictly enforcing downward monotonicity and validating feature sets through architectural safeguards, we ensure that the compiler remains both fast and correct, providing immediate value to developers while scaling to meet the future demands of web development.

\end{document}